\pdfoutput=1
\documentclass[11pt,twoside,a4paper,pdftex,cmspaper,final,collab]{cms-tdr}

\begin{document}\cmsNoteHeader{CFT-09-022}
%%%%%%%%%%%%%%%%%%%%%%%%%%%%%%%%%%%%%%%%%%%%%%%%%%%%%%%%%%%%%%%%%%%%
%
%  Common definitions
%
%  N.B. use of \providecommand rather than \newcommand means
%       that a definition is ignored if already specified
%
%                                              L. Taylor 18 Feb 2005
%%%%%%%%%%%%%%%%%%%%%%%%%%%%%%%%%%%%%%%%%%%%%%%%%%%%%%%%%%%%%%%%%%%%

%%%%%%%%%%%%%%%%%%%%%%%%%%%%%%%%%%%%%%%%%%%%%%%%%%%%%%%%%%%%%%%%%%%%
%
% Hyphenations (only need to add here if you get a nasty word break)
%
\hyphenation{env-iron-men-tal}%    just an example
\hyphenation{had-ron-i-za-tion}
\hyphenation{cal-or-i-me-ter}
\hyphenation{de-vices}
%
% Hyphenations-end
%
%CVS info. These are modified by cvs at checkout time.
% The l st version of these macros found before the maketitle will be the one on the front page,
%  so o nly the main file is tracked.
% Edit by hand with  care!
\RCS$Revision: 1.22 $
\RCS$Date: 2009/12/16 10:06:37 $
\RCS$ Name:  $
%%%%%%%%%%%%% ptdr definitions %%%%%%%%%%%%%%%%%%%%%
%%%%%%%%%%%%%%%%%%%%%%%%%%%%%%%%%%%%%%%%%%%%%%%%%%%%%%%%%%%%%%%%%%%%
%
%  Common definitions
%
%  N.B. use of \providecommand rather than \newcommand means
%       that a definition is ignored if already specified
%
%                                              L. Taylor 18 Feb 2005
%%%%%%%%%%%%%%%%%%%%%%%%%%%%%%%%%%%%%%%%%%%%%%%%%%%%%%%%%%%%%%%%%%%%

% Some shorthand
% turn off italics
\newcommand {\etal}{\mbox{et al.}\xspace} %et al. - no preceding comma
\newcommand {\ie}{\mbox{i.e.}\xspace}     %i.e.
\newcommand {\eg}{\mbox{e.g.}\xspace}     %e.g.
\newcommand {\etc}{\mbox{etc.}\xspace}     %etc.
\newcommand {\vs}{\mbox{\sl vs.}\xspace}      %vs.
\newcommand {\mdash}{\ensuremath{\mathrm{-}}} % for use within formulas

% some terms whose definition we may change
\newcommand {\Lone}{Level-1\xspace} % Level-1 or L1 ?
\newcommand {\Ltwo}{Level-2\xspace}
\newcommand {\Lthree}{Level-3\xspace}

% Some software programs (alphabetized)
\providecommand{\ACERMC} {\textsc{AcerMC}\xspace}
\providecommand{\ALPGEN} {{\textsc{alpgen}}\xspace}
\providecommand{\CHARYBDIS} {{\textsc{charybdis}}\xspace}
\providecommand{\CMKIN} {\textsc{cmkin}\xspace}
\providecommand{\CMSIM} {{\textsc{cmsim}}\xspace}
\providecommand{\CMSSW} {{\textsc{cmssw}}\xspace}
\providecommand{\COBRA} {{\textsc{cobra}}\xspace}
\providecommand{\COCOA} {{\textsc{cocoa}}\xspace}
\providecommand{\COMPHEP} {\textsc{CompHEP}\xspace}
\providecommand{\EVTGEN} {{\textsc{evtgen}}\xspace}
\providecommand{\FAMOS} {{\textsc{famos}}\xspace}
\providecommand{\GARCON} {\textsc{garcon}\xspace}
\providecommand{\GARFIELD} {{\textsc{garfield}}\xspace}
\providecommand{\GEANE} {{\textsc{geane}}\xspace}
\providecommand{\GEANTfour} {{\textsc{geant4}}\xspace}
\providecommand{\GEANTthree} {{\textsc{geant3}}\xspace}
\providecommand{\GEANT} {{\textsc{geant}}\xspace}
\providecommand{\HDECAY} {\textsc{hdecay}\xspace}
\providecommand{\HERWIG} {{\textsc{herwig}}\xspace}
\providecommand{\HIGLU} {{\textsc{higlu}}\xspace}
\providecommand{\HIJING} {{\textsc{hijing}}\xspace}
\providecommand{\IGUANA} {\textsc{iguana}\xspace}
\providecommand{\ISAJET} {{\textsc{isajet}}\xspace}
\providecommand{\ISAPYTHIA} {{\textsc{isapythia}}\xspace}
\providecommand{\ISASUGRA} {{\textsc{isasugra}}\xspace}
\providecommand{\ISASUSY} {{\textsc{isasusy}}\xspace}
\providecommand{\ISAWIG} {{\textsc{isawig}}\xspace}
\providecommand{\MADGRAPH} {\textsc{MadGraph}\xspace}
\providecommand{\MCATNLO} {\textsc{mc@nlo}\xspace}
\providecommand{\MCFM} {\textsc{mcfm}\xspace}
\providecommand{\MILLEPEDE} {{\textsc{millepede}}\xspace}
\providecommand{\ORCA} {{\textsc{orca}}\xspace}
\providecommand{\OSCAR} {{\textsc{oscar}}\xspace}
\providecommand{\PHOTOS} {\textsc{photos}\xspace}
\providecommand{\PROSPINO} {\textsc{prospino}\xspace}
\providecommand{\PYTHIA} {{\textsc{pythia}}\xspace}
\providecommand{\SHERPA} {{\textsc{sherpa}}\xspace}
\providecommand{\TAUOLA} {\textsc{tauola}\xspace}
\providecommand{\TOPREX} {\textsc{TopReX}\xspace}
\providecommand{\XDAQ} {{\textsc{xdaq}}\xspace}

%  Experiments
\newcommand {\DZERO}{D\O\xspace}     %etc.

% Measurements and units...

\newcommand{\de}{\ensuremath{^\circ}}
\newcommand{\ten}[1]{\ensuremath{\times \text{10}^\text{#1}}}
\newcommand{\unit}[1]{\ensuremath{\text{\,#1}}\xspace}
\newcommand{\mum}{\ensuremath{\,\mu\text{m}}\xspace}
\newcommand{\micron}{\ensuremath{\,\mu\text{m}}\xspace}
\newcommand{\cm}{\ensuremath{\,\text{cm}}\xspace}
\newcommand{\mm}{\ensuremath{\,\text{mm}}\xspace}
\newcommand{\mus}{\ensuremath{\,\mu\text{s}}\xspace}
\newcommand{\keV}{\ensuremath{\,\text{ke\hspace{-.08em}V}}\xspace}
\newcommand{\MeV}{\ensuremath{\,\text{Me\hspace{-.08em}V}}\xspace}
\newcommand{\GeV}{\ensuremath{\,\text{Ge\hspace{-.08em}V}}\xspace}
\newcommand{\TeV}{\ensuremath{\,\text{Te\hspace{-.08em}V}}\xspace}
\newcommand{\PeV}{\ensuremath{\,\text{Pe\hspace{-.08em}V}}\xspace}
\newcommand{\keVc}{\ensuremath{{\,\text{ke\hspace{-.08em}V\hspace{-0.16em}/\hspace{-0.08em}c}}}\xspace}
\newcommand{\MeVc}{\ensuremath{{\,\text{Me\hspace{-.08em}V\hspace{-0.16em}/\hspace{-0.08em}c}}}\xspace}
\newcommand{\GeVc}{\ensuremath{{\,\text{Ge\hspace{-.08em}V\hspace{-0.16em}/\hspace{-0.08em}c}}}\xspace}
\newcommand{\TeVc}{\ensuremath{{\,\text{Te\hspace{-.08em}V\hspace{-0.16em}/\hspace{-0.08em}c}}}\xspace}
\newcommand{\keVcc}{\ensuremath{{\,\text{ke\hspace{-.08em}V\hspace{-0.16em}/\hspace{-0.08em}c}^\text{2}}}\xspace}
\newcommand{\MeVcc}{\ensuremath{{\,\text{Me\hspace{-.08em}V\hspace{-0.16em}/\hspace{-0.08em}c}^\text{2}}}\xspace}
\newcommand{\GeVcc}{\ensuremath{{\,\text{Ge\hspace{-.08em}V\hspace{-0.16em}/\hspace{-0.08em}c}^\text{2}}}\xspace}
\newcommand{\TeVcc}{\ensuremath{{\,\text{Te\hspace{-.08em}V\hspace{-0.16em}/\hspace{-0.08em}c}^\text{2}}}\xspace}

\newcommand{\pbinv} {\mbox{\ensuremath{\,\text{pb}^\text{$-$1}}}\xspace}
\newcommand{\fbinv} {\mbox{\ensuremath{\,\text{fb}^\text{$-$1}}}\xspace}
\newcommand{\nbinv} {\mbox{\ensuremath{\,\text{nb}^\text{$-$1}}}\xspace}
\newcommand{\percms}{\ensuremath{\,\text{cm}^\text{$-$2}\,\text{s}^\text{$-$1}}\xspace}
\newcommand{\lumi}{\ensuremath{\mathcal{L}}\xspace}
\newcommand{\Lumi}{\ensuremath{\mathcal{L}}\xspace}%both upper and lower
%
% Need a convention here:
\newcommand{\LvLow}  {\ensuremath{\mathcal{L}=\text{10}^\text{32}\,\text{cm}^\text{$-$2}\,\text{s}^\text{$-$1}}\xspace}
\newcommand{\LLow}   {\ensuremath{\mathcal{L}=\text{10}^\text{33}\,\text{cm}^\text{$-$2}\,\text{s}^\text{$-$1}}\xspace}
\newcommand{\lowlumi}{\ensuremath{\mathcal{L}=\text{2}\times \text{10}^\text{33}\,\text{cm}^\text{$-$2}\,\text{s}^\text{$-$1}}\xspace}
\newcommand{\LMed}   {\ensuremath{\mathcal{L}=\text{2}\times \text{10}^\text{33}\,\text{cm}^\text{$-$2}\,\text{s}^\text{$-$1}}\xspace}
\newcommand{\LHigh}  {\ensuremath{\mathcal{L}=\text{10}^\text{34}\,\text{cm}^\text{$-$2}\,\text{s}^\text{$-$1}}\xspace}
\newcommand{\hilumi} {\ensuremath{\mathcal{L}=\text{10}^\text{34}\,\text{cm}^\text{$-$2}\,\text{s}^\text{$-$1}}\xspace}

% Some usual physics terms

\newcommand{\zp}{\ensuremath{\mathrm{Z}^\prime}\xspace}

% SM (still to be classified)

\newcommand{\kt}{\ensuremath{k_{\mathrm{T}}}\xspace}
\newcommand{\BC}{\ensuremath{{B_{\mathrm{c}}}}\xspace}
\newcommand{\bbarc}{\ensuremath{{\overline{b}c}}\xspace}
\newcommand{\bbbar}{\ensuremath{{b\overline{b}}}\xspace}
\newcommand{\ccbar}{\ensuremath{{c\overline{c}}}\xspace}
\newcommand{\JPsi}{\ensuremath{{J}/\psi}\xspace}
\newcommand{\bspsiphi}{\ensuremath{B_s \to \JPsi\, \phi}\xspace}
\newcommand{\AFB}{\ensuremath{A_\mathrm{FB}}\xspace}
\newcommand{\EE}{\ensuremath{e^+e^-}\xspace}
\newcommand{\MM}{\ensuremath{\mu^+\mu^-}\xspace}
\newcommand{\TT}{\ensuremath{\tau^+\tau^-}\xspace}
\newcommand{\wangle}{\ensuremath{\sin^{2}\theta_{\mathrm{eff}}^\mathrm{lept}(M^2_\mathrm{Z})}\xspace}
\newcommand{\ttbar}{\ensuremath{{t\overline{t}}}\xspace}
\newcommand{\stat}{\ensuremath{\,\text{(stat.)}}\xspace}
\newcommand{\syst}{\ensuremath{\,\text{(syst.)}}\xspace}
% these moved to similar defs
%\newcommand{\Etmiss}{\ensuremath{E_{\mathrm{T}\!{\rm miss}}}}
%\newcommand{\VEtmiss}{\ensuremath{{\vec E}_{\mathrm{T}\!{\rm miss}}}}

%%%  E-gamma definitions
\newcommand{\HGG}{\ensuremath{\mathrm{H}\to\gamma\gamma}}
\newcommand{\gev}{\GeV}
\newcommand{\GAMJET}{\ensuremath{\gamma + \mathrm{jet}}}
\newcommand{\PPTOJETS}{\ensuremath{\mathrm{pp}\to\mathrm{jets}}}
\newcommand{\PPTOGG}{\ensuremath{\mathrm{pp}\to\gamma\gamma}}
\newcommand{\PPTOGAMJET}{\ensuremath{\mathrm{pp}\to\gamma +
\mathrm{jet}
}}
\newcommand{\MH}{\ensuremath{\mathrm{M_{\mathrm{H}}}}}
\newcommand{\RNINE}{\ensuremath{\mathrm{R}_\mathrm{9}}}
\newcommand{\DR}{\ensuremath{\Delta\mathrm{R}}}

% Physics symbols ...

\newcommand{\PT}{\ensuremath{p_{\mathrm{T}}}\xspace}
\newcommand{\pt}{\ensuremath{p_{\mathrm{T}}}\xspace}
\newcommand{\ET}{\ensuremath{E_{\mathrm{T}}}\xspace}
\newcommand{\HT}{\ensuremath{H_{\mathrm{T}}}\xspace}
\newcommand{\et}{\ensuremath{E_{\mathrm{T}}}\xspace}
\newcommand{\Em}{\ensuremath{E\!\!\!/}\xspace}
\newcommand{\Pm}{\ensuremath{p\!\!\!/}\xspace}
\newcommand{\PTm}{\ensuremath{{p\!\!\!/}_{\mathrm{T}}}\xspace}
\newcommand{\ETm}{\ensuremath{E_{\mathrm{T}}^{\mathrm{miss}}}\xspace}
\newcommand{\MET}{\ensuremath{E_{\mathrm{T}}^{\mathrm{miss}}}\xspace}
\newcommand{\ETmiss}{\ensuremath{E_{\mathrm{T}}^{\mathrm{miss}}}\xspace}
\newcommand{\VEtmiss}{\ensuremath{{\vec E}_{\mathrm{T}}^{\mathrm{miss}}}\xspace}

%%%%%%
% From Albert
%

\newcommand{\ga}{\ensuremath{\gtrsim}}
\newcommand{\la}{\ensuremath{\lesssim}}
\newcommand{\swsq}{\ensuremath{\sin^2\theta_W}\xspace}
\newcommand{\cwsq}{\ensuremath{\cos^2\theta_W}\xspace}
\newcommand{\tanb}{\ensuremath{\tan\beta}\xspace}
\newcommand{\tanbsq}{\ensuremath{\tan^{2}\beta}\xspace}
\newcommand{\sidb}{\ensuremath{\sin 2\beta}\xspace}
\newcommand{\alpS}{\ensuremath{\alpha_S}\xspace}
\newcommand{\alpt}{\ensuremath{\tilde{\alpha}}\xspace}

\newcommand{\QL}{\ensuremath{Q_L}\xspace}
\newcommand{\sQ}{\ensuremath{\tilde{Q}}\xspace}
\newcommand{\sQL}{\ensuremath{\tilde{Q}_L}\xspace}
\newcommand{\ULC}{\ensuremath{U_L^C}\xspace}
\newcommand{\sUC}{\ensuremath{\tilde{U}^C}\xspace}
\newcommand{\sULC}{\ensuremath{\tilde{U}_L^C}\xspace}
\newcommand{\DLC}{\ensuremath{D_L^C}\xspace}
\newcommand{\sDC}{\ensuremath{\tilde{D}^C}\xspace}
\newcommand{\sDLC}{\ensuremath{\tilde{D}_L^C}\xspace}
\newcommand{\LL}{\ensuremath{L_L}\xspace}
\newcommand{\sL}{\ensuremath{\tilde{L}}\xspace}
\newcommand{\sLL}{\ensuremath{\tilde{L}_L}\xspace}
\newcommand{\ELC}{\ensuremath{E_L^C}\xspace}
\newcommand{\sEC}{\ensuremath{\tilde{E}^C}\xspace}
\newcommand{\sELC}{\ensuremath{\tilde{E}_L^C}\xspace}
\newcommand{\sEL}{\ensuremath{\tilde{E}_L}\xspace}
\newcommand{\sER}{\ensuremath{\tilde{E}_R}\xspace}
\newcommand{\sFer}{\ensuremath{\tilde{f}}\xspace}
\newcommand{\sQua}{\ensuremath{\tilde{q}}\xspace}
\newcommand{\sUp}{\ensuremath{\tilde{u}}\xspace}
\newcommand{\suL}{\ensuremath{\tilde{u}_L}\xspace}
\newcommand{\suR}{\ensuremath{\tilde{u}_R}\xspace}
\newcommand{\sDw}{\ensuremath{\tilde{d}}\xspace}
\newcommand{\sdL}{\ensuremath{\tilde{d}_L}\xspace}
\newcommand{\sdR}{\ensuremath{\tilde{d}_R}\xspace}
\newcommand{\sTop}{\ensuremath{\tilde{t}}\xspace}
\newcommand{\stL}{\ensuremath{\tilde{t}_L}\xspace}
\newcommand{\stR}{\ensuremath{\tilde{t}_R}\xspace}
\newcommand{\stone}{\ensuremath{\tilde{t}_1}\xspace}
\newcommand{\sttwo}{\ensuremath{\tilde{t}_2}\xspace}
\newcommand{\sBot}{\ensuremath{\tilde{b}}\xspace}
\newcommand{\sbL}{\ensuremath{\tilde{b}_L}\xspace}
\newcommand{\sbR}{\ensuremath{\tilde{b}_R}\xspace}
\newcommand{\sbone}{\ensuremath{\tilde{b}_1}\xspace}
\newcommand{\sbtwo}{\ensuremath{\tilde{b}_2}\xspace}
\newcommand{\sLep}{\ensuremath{\tilde{l}}\xspace}
\newcommand{\sLepC}{\ensuremath{\tilde{l}^C}\xspace}
\newcommand{\sEl}{\ensuremath{\tilde{e}}\xspace}
\newcommand{\sElC}{\ensuremath{\tilde{e}^C}\xspace}
\newcommand{\seL}{\ensuremath{\tilde{e}_L}\xspace}
\newcommand{\seR}{\ensuremath{\tilde{e}_R}\xspace}
\newcommand{\snL}{\ensuremath{\tilde{\nu}_L}\xspace}
\newcommand{\sMu}{\ensuremath{\tilde{\mu}}\xspace}
\newcommand{\sNu}{\ensuremath{\tilde{\nu}}\xspace}
\newcommand{\sTau}{\ensuremath{\tilde{\tau}}\xspace}
\newcommand{\Glu}{\ensuremath{g}\xspace}
\newcommand{\sGlu}{\ensuremath{\tilde{g}}\xspace}
\newcommand{\Wpm}{\ensuremath{W^{\pm}}\xspace}
\newcommand{\sWpm}{\ensuremath{\tilde{W}^{\pm}}\xspace}
\newcommand{\Wz}{\ensuremath{W^{0}}\xspace}
\newcommand{\sWz}{\ensuremath{\tilde{W}^{0}}\xspace}
\newcommand{\sWino}{\ensuremath{\tilde{W}}\xspace}
\newcommand{\Bz}{\ensuremath{B^{0}}\xspace}
\newcommand{\sBz}{\ensuremath{\tilde{B}^{0}}\xspace}
\newcommand{\sBino}{\ensuremath{\tilde{B}}\xspace}
\newcommand{\Zz}{\ensuremath{Z^{0}}\xspace}
\newcommand{\sZino}{\ensuremath{\tilde{Z}^{0}}\xspace}
\newcommand{\sGam}{\ensuremath{\tilde{\gamma}}\xspace}
\newcommand{\chiz}{\ensuremath{\tilde{\chi}^{0}}\xspace}
\newcommand{\chip}{\ensuremath{\tilde{\chi}^{+}}\xspace}
\newcommand{\chim}{\ensuremath{\tilde{\chi}^{-}}\xspace}
\newcommand{\chipm}{\ensuremath{\tilde{\chi}^{\pm}}\xspace}
\newcommand{\Hone}{\ensuremath{H_{d}}\xspace}
\newcommand{\sHone}{\ensuremath{\tilde{H}_{d}}\xspace}
\newcommand{\Htwo}{\ensuremath{H_{u}}\xspace}
\newcommand{\sHtwo}{\ensuremath{\tilde{H}_{u}}\xspace}
\newcommand{\sHig}{\ensuremath{\tilde{H}}\xspace}
\newcommand{\sHa}{\ensuremath{\tilde{H}_{a}}\xspace}
\newcommand{\sHb}{\ensuremath{\tilde{H}_{b}}\xspace}
\newcommand{\sHpm}{\ensuremath{\tilde{H}^{\pm}}\xspace}
\newcommand{\hz}{\ensuremath{h^{0}}\xspace}
\newcommand{\Hz}{\ensuremath{H^{0}}\xspace}
\newcommand{\Az}{\ensuremath{A^{0}}\xspace}
\newcommand{\Hpm}{\ensuremath{H^{\pm}}\xspace}
\newcommand{\sGra}{\ensuremath{\tilde{G}}\xspace}
\newcommand{\mtil}{\ensuremath{\tilde{m}}\xspace}
\newcommand{\rpv}{\ensuremath{\rlap{\kern.2em/}R}\xspace}
\newcommand{\LLE}{\ensuremath{LL\bar{E}}\xspace}
\newcommand{\LQD}{\ensuremath{LQ\bar{D}}\xspace}
\newcommand{\UDD}{\ensuremath{\overline{UDD}}\xspace}
\newcommand{\Lam}{\ensuremath{\lambda}\xspace}
\newcommand{\Lamp}{\ensuremath{\lambda'}\xspace}
\newcommand{\Lampp}{\ensuremath{\lambda''}\xspace}
\newcommand{\spinbd}[2]{\ensuremath{\bar{#1}_{\dot{#2}}}\xspace}

\newcommand{\MD}{\ensuremath{{M_\mathrm{D}}}\xspace}% ED mass
\newcommand{\Mpl}{\ensuremath{{M_\mathrm{Pl}}}\xspace}% Planck mass
\newcommand{\Rinv} {\ensuremath{{R}^{-1}}\xspace}

%%%%%%%%%%%%%%%%%%%%%%%%%%%%%%%%%%%%%%%%%%%%%%%%%%%%%%%%%%%%%%%%%%%%
%
% Hyphenations (only need to add here if you get a nasty word break)
%
\hyphenation{en-viron-men-tal}%    just an example

%%%%%%%%%%%%%%%  Title page %%%%%%%%%%%%%%%%%%%%%%%%
\cmsNoteHeader{09-022}
\title{Performance of the CMS Drift-Tube Chamber Local Trigger with Cosmic Rays}% Force line breaks with \\

%Author is always "The CMS Collaboration" for PAS, so author, etc will be ignored
%\address[cern]{CERN}
%\address[neu]{Northeastern University}
%\author[neu]{George Alverson}\author[neu]{Lucas Taylor}\author[cern]{A. Cern Person}

% please supply the date in yyyy/mm/dd format. Today has been
% redefined to do so, but it should be fixed as of the final release date.
\date{\today}

% note that you cannot use \verb in the abstract text
\abstract{
The performance of the Local Trigger based on the drift-tube
system of the CMS experiment has been studied using muons from 
cosmic ray events collected during the commissioning of the detector in 2008.
The properties of the system are extensively tested and compared with the simulation.
The effect of the random arrival time of the cosmic rays on the
trigger performance is reported, and 
the results are compared with the design expectations for proton-proton collisions 
and with previous measurements obtained with muon beams.
}

% these need to be filled in by hand and should (MUST) match the info
% in the TeX equivalents less the TeX markup
\hypersetup{%
%pdfauthor={George Alverson,Lucas Taylor, A. Cern Person},%
pdfauthor={Stefano Marcellini - INFN Bologna, Italy},%
pdftitle={Performance of the CMS Drift-Tube Local Trigger with Cosmic Rays},%
pdfsubject={CMS},%
pdfkeywords={CMS, CRAFT, trigger, muon}}

\maketitle %maketitle comes after all the front information has been supplied

%%%%%%%%%%%%%%%%%%%%%%%%%%%%%%%%  Begin text %%%%%%%%%%%%%%%%%%%%%%%%%%%%%
\section{Introduction}
The primary goal of the Compact Muon Solenoid (CMS) experiment~\cite{cms_det_paper} 
is to explore particle physics at the TeV energy scale 
exploiting the proton-proton collisions delivered by the Large Hadron Collider (LHC)~\cite{LHC}.
During October-November 2008 CMS conducted a month-long data taking exercise
with the goal of commissioning the experiment for extended operation~\cite{CRAFTGeneral}.
The experiment 
recorded 270 million
cosmic ray triggered events with the solenoid at its nominal axial field
strength of 3.8~T.
Prior to this, and during the final installation phase of the experiment, a
series of week-long commissioning exercises to record cosmic ray events took
place from May to September 2008, without magnetic field. Over 300 million
cosmic ray triggers were accumulated during this data taking period.
The drift-tube muon system of the detector continuously provided the trigger to the experiment and enabled 
muon track reconstruction. The drift-tube Local Trigger is the component of the 
trigger system that provides the information to
reconstruct muon trigger candidates in the drift-tube system.

This paper describes the performance of the drift-tube Local Trigger system
determined from data collected in this data taking period. 
Results include measurements of the system efficiency, 
the performance of the bunch crossing identification, 
the position and angular resolution of the muon-trigger candidate, and the 
false dimuon trigger rate. The trigger data are also compared with
software emulation. Such studies test the reliability of the system
and improve the understanding of the muon trigger performance
in preparation for LHC operation. 
The trigger system design is optimized for muons 
produced in bunched-beam collisions.
The random arrival time of the cosmic rays degrades the average performance 
compared to the expectations for proton-proton collisions.
However, results for particles traversing the detector with the proper timing
show that the system will perform as required for muons produced at the LHC.

\section{The Experimental Apparatus}
\label{dttrigger}

%\subsection{The CMS Experiment and the Drift-Tube Muon System}

The central feature of the CMS apparatus is a
superconducting solenoid, of $\rm 6\,m$ internal diameter, providing a field of
$\rm 3.8\,T$. Within the field volume are the silicon pixel and strip tracker, the
crystal electromagnetic calorimeter (ECAL), and the brass/scintillator hadron
calorimeter (HCAL).  Muons are measured in gas-ionization detectors embedded
in the steel return yoke.  In addition to the barrel and endcap detectors,
CMS has extensive forward calorimetry.
A detailed description of the CMS experiment
can be found in Ref.~\cite{cms_det_paper}.

CMS uses a right-handed coordinate system, with the origin at the
nominal collision point, the $x$-axis pointing to the center of the
LHC, the $y$-axis pointing up (perpendicular to the LHC plane), and
the $z$-axis along the anticlockwise-beam direction.
The polar angle, $\theta$, is measured from the positive $z$-axis and
the azimuthal angle, $\phi$, is measured in the $x$-$y$ plane.
 
The CMS experiment has a two-level trigger system: the Level-1 Trigger and the High Level Trigger.
The Level-1 Trigger is composed of custom
hardware processors and it uses information from the calorimeters and muon
detectors to select the events with a constant latency of about 3.5 $\mu$s. The High Level Trigger
processor farm further decreases the event rate from a maximum input rate of  $\rm 100\,kHz$ to an event recording rate of approximately
$\rm 100\,Hz$~\cite{trig_TDR,workflow}.

 \begin{figure}[hbtp]
\includegraphics[width=16.4cm]{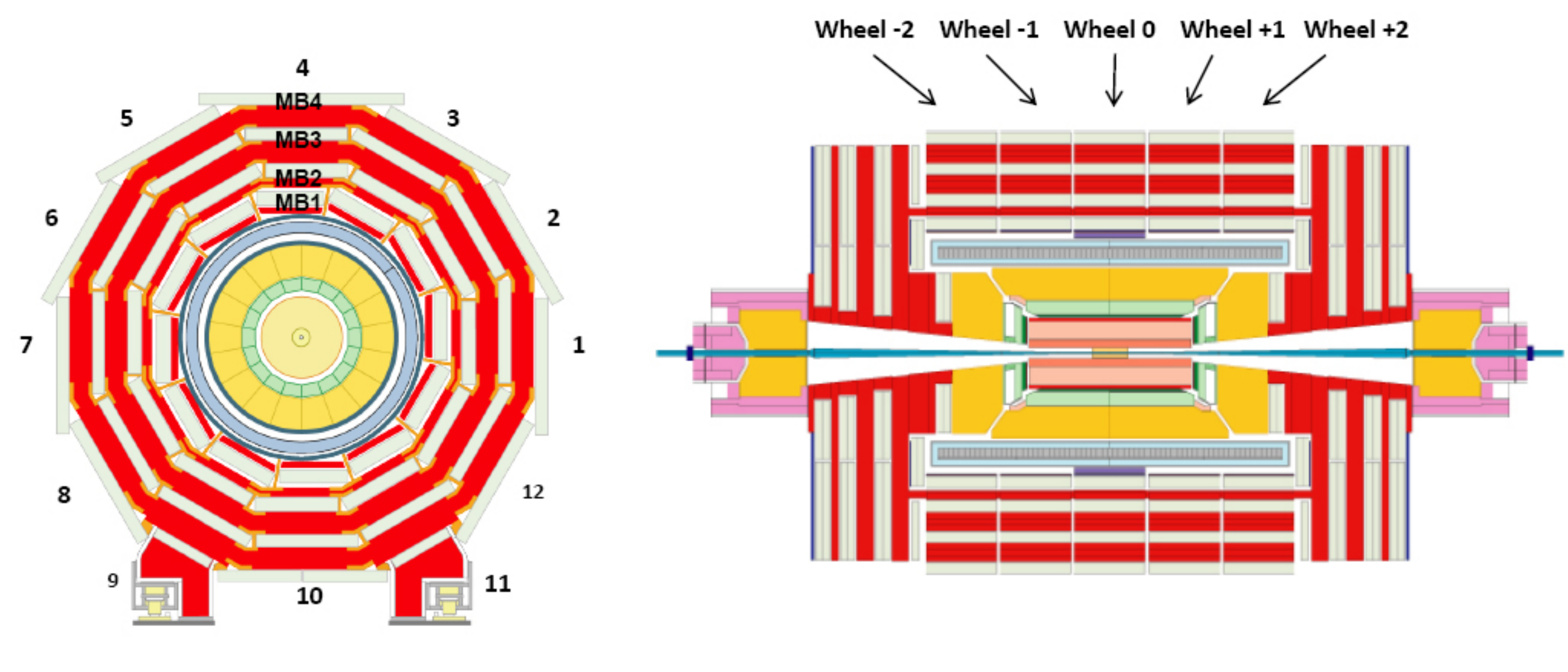}
    \caption{Left: schematic view of one of the five  wheels of the muon barrel system of the CMS 
experiment, in the $x$-$y$ plane.
Right: the five wheels are visible in the 
longitudinal view of the CMS detector.}
    \label{fig:figure1}
%  \end{center}
\end{figure}
The muon detection system covers the pseudorapidity range $|\eta|< 2.4$
with detection planes based on three different technologies: drift tubes (DT), cathode
strip chambers (CSC), and resistive plate chambers (RPC)~\cite{muon_tdr}.
The DT system is located in the barrel part of the detector, extending over the range $|\eta|< 1.3$,
whereas CSCs are used in the endcaps, in the region $0.8 < |\eta|< 2.4$.
The RPC system covers the region $|\eta|< 1.6$.
Combining the measured tracks from charged
particles in the muon system with those
measured in the silicon tracker results in a transverse
momentum resolution between 1 and $\rm 5\,\%$ for transverse momenta up to
$1\, \rm ~TeV$/$c$~\cite{phys_tdr}.
The muon detector also
provides a highly efficient muon trigger.
The usage of RPCs and DTs in the barrel, and RPCs and CSCs in the endcaps, 
makes the system robust and redundant. 
Both the DT and CSC detectors
perform muon tracking 
and participate in the Level-1 muon trigger, independently of the RPCs.

\subsection{The Drift-Tube Muon System}

The DT system consists of 240 muon chambers.
Figure~\ref{fig:figure1} shows
how the muon chambers are arranged in 12 sectors with four station types
in each of the five wheels of the CMS barrel. 
The four station types are called MB1, MB2, MB3, and MB4 from inside to outside, 
where MB stands for Muon Barrel.
Although the chambers of station type MB4 in sectors 4 and 10 are 
physically divided in two parts, in this paper they are considered as a single unit.
In each chamber eight layers of DT cells with wires parallel to 
the beam direction are arranged in two "superlayers" of four  
layers each, devoted to trigger and position measurements in 
the $x$-$y$ plane. With the exception of type MB4, all chambers are 
equipped with an extra "superlayer" of four layers to 
perform measurements along the $z$ coordinate using DT 
cells with wires orthogonal to the beam direction.

Chambers are equipped with Time to Digital Converter (TDC) units, 
which measure the drift time in each cell, 
as well as with dedicated electronics to 
perform the Level-1 trigger algorithm. For tracking purposes, 
the signals recorded by the TDC 
units are converted to position coordinates (called ``hits'') using the
space-time relation determined with the knowledge of the drift velocity. 
In each chamber, track segments are reconstructed using a linear fit to 
the hits, and are used for 
the offline reconstruction of the muon tracks~\cite{localtrack}.
For triggering purposes, the DT Local Trigger 
algorithm searches for aligned hits, called 
trigger segments, in each chamber, using dedicated electronics and 
algorithms that are independent of the TDC track fitting.

\subsection{The Drift-Tube Local Trigger}

The DT Local Trigger system and its performance are described in 
detail in Refs.~\cite{trig_TDR, NIM_2004, NIM_2007}. Dedicated adjustments to the configuration of the 
DT trigger were needed to provide high efficiency for cosmic ray muons, which have a
different origin, direction, and timing compared to muons
from proton-proton collisions~\cite{CRAFT-L1}.
In this section, only the main functionalities and characteristics are
summarized. 
The trigger segments are found separately in the transverse plane $x$-$y$ 
(called $\phi$ view) and in the plane that contains the $z$ direction (called $\theta$ view).

The maximum drift time in the DT 
system is almost $400 \,\rm{ns}$, 
which is much longer than the $25\,\rm{ns}$ interval that separates two consecutive collisions. 
Therefore, the trigger system must first associate each trigger segment to the correct bunch crossing 
(BX).
For each BX the trigger system provides up to two trigger segments per chamber in the $\phi$ view, and one 
in the $\theta$ view. In the $\phi$ view, each trigger segment is associated with the following quantities: the BX, 
at which the corresponding muon candidate was produced;
the position and direction in the local coordinate system of the chamber; a quality word 
describing how many aligned DT hits were found; and a bit flagging the segment as a
first or second candidate, ordered according to their assigned quality, for that BX. 
One set of such quantities is called a ``trigger primitive''.

\vspace{-1.cm}
\begin{figure}[hbtp]
  \begin{center}
\includegraphics[width=10cm]{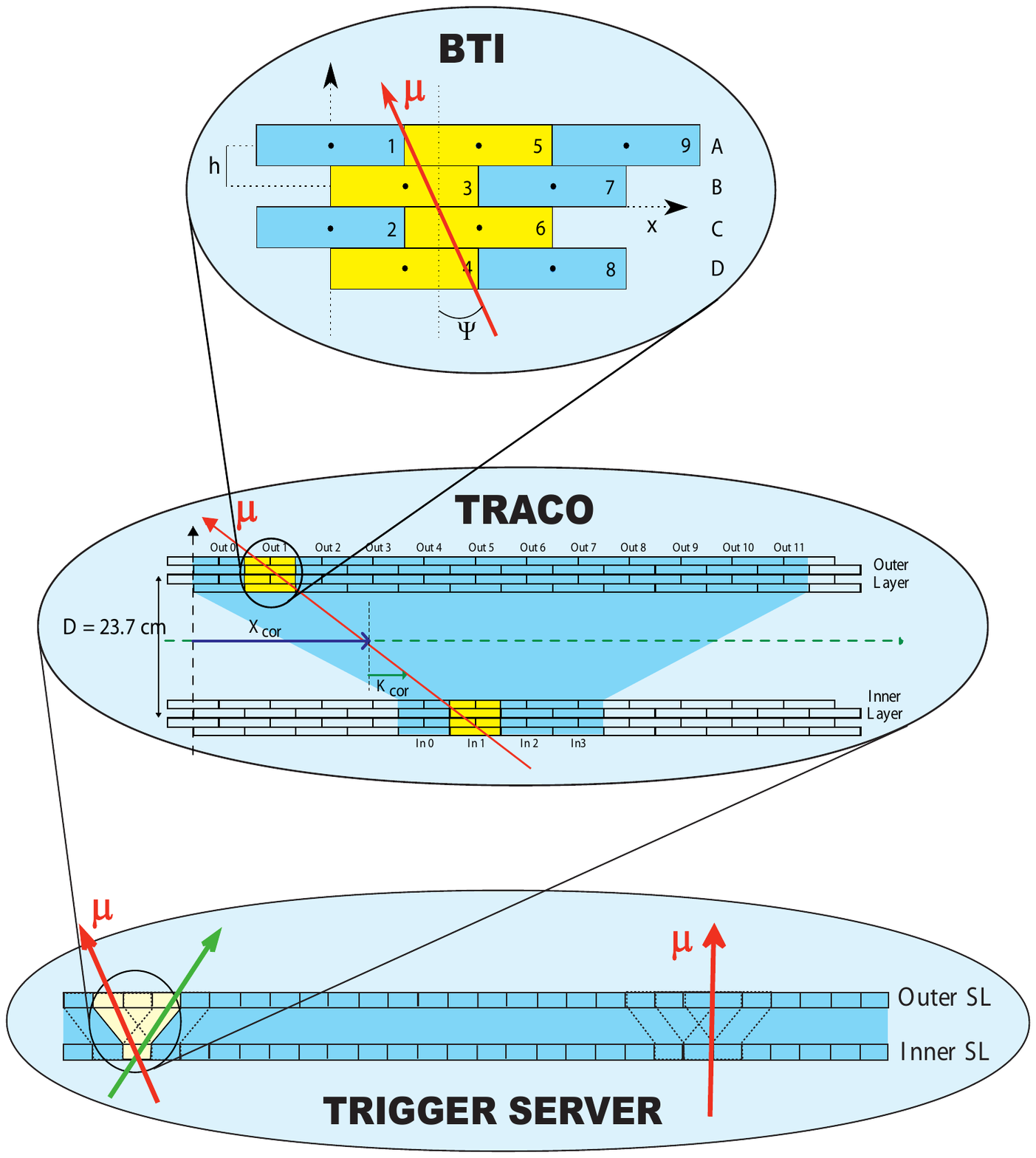}
\vspace{-2.cm}
    \caption{The main components of the DT Local Trigger in the $\phi$ view of a muon chamber. 
The BTIs detect hit alignments within each superlayer,
the TRACOs search for a proper matching between superlayers, and the Trigger Server 
selects the best two candidates in the chamber and applies a ghost-suppression algorithm.  }
    \label{fig:figure2}
  \end{center}
\end{figure}

Trigger segments 
in the $\theta$ view are issued if there are three or four aligned 
hits in the corresponding superlayer. 
The $\theta$ trigger primitive includes the assigned BX, a word defining the location in the chamber, and 
its quality word.
During 
LHC operation, the system in the $\theta$ view will be configured to accept only trigger segments 
pointing to the interaction region. This condition was not applied for data taking with cosmic rays.

The trigger algorithm that provides the trigger primitives in the $\phi$ view
works in three logical steps that are sketched in Fig.~\ref{fig:figure2}.
For every BX, the Bunch and Track Identifiers (BTIs) 
search for track segment candidates made of three or four aligned DT hits within a superlayer. 
Each candidate is
processed by the Track Correlators (TRACOs), which search for a
proper angular matching 
between the candidates from the two superlayers in each chamber. 
TRACO candidates are then sent to the Trigger Server, 
which performs a ghost-suppression mechanism and then selects for each BX 
the two trigger primitives with highest quality.
The quality is defined according 
to the number of aligned hits in the trigger segment.
An alignment of hits in a 
superlayer is labeled ``High'' (H) or ``Low'' (L) if it includes
four or three DT hits, respectively. The quality of a trigger segment in the $\phi$ view is labeled 
HH, HL, or LL if a matching (correlation), both in angle and BX, is found between 
the segments of the two superlayers. In this case the trigger segments are called ``correlated''.
If no correlation is found,
and four aligned hits are detected in the inner (outer) superlayer of the chamber 
(where inner and outer are defined according to their distance from the center of the CMS detector), 
the segment quality is 
labeled as $\rm HI$ ($\rm HO$).

The DT Local trigger is designed 
for triggering on particles produced 
synchronously with the bunch crossing signal. 
One of the effects of the random 
arrival time of the cosmic muons 
is the production of spurious uncorrelated
trigger segments with wrong bunch crossing assignment. 
Therefore, 
to reduce the rate of such spurious triggers, 
the DT Local Trigger was configured such that the presence of
a $\theta$ trigger segment was required to 
accept uncorrelated H-quality 
triggers in the $\phi$ view.  
At the LHC, 
where muons are synchronously produced with 
the bunch crossing signal, 
spurious triggers will be largely suppressed, so that
the trigger configuration will be relaxed 
to accept also uncorrelated L-quality triggers.

%For data taking with cosmic rays, the DT Local Trigger was configured such that the
%presence of a $\theta$ trigger segment was required to accept uncorrelated H-quality triggers in the $\phi$ view.
%This requirement reduces the rate of spurious uncorrelated trigger segments 
%that originate from the random arrival time of the cosmic ray muons, and it will be relaxed for triggering at the LHC.

The ghost-suppression 
mechanism is performed %by the Trigger Server 
to discard additional trigger candidates
attributable to false signals (``ghosts'').  
The false trigger segments mostly arise from the fact that adjacent TRACO units share a common group of DT cells, 
and a hit alignment can be found twice.
The segment selection and ghost-suppression algorithms are flexible and can 
be configured to match various experimental conditions, such as the presence of groups of noisy or
disconnected DT cells in a chamber.
   
\begin{figure}[hbtp]
  \begin{center}
    \includegraphics[width=8cm,height=8cm]{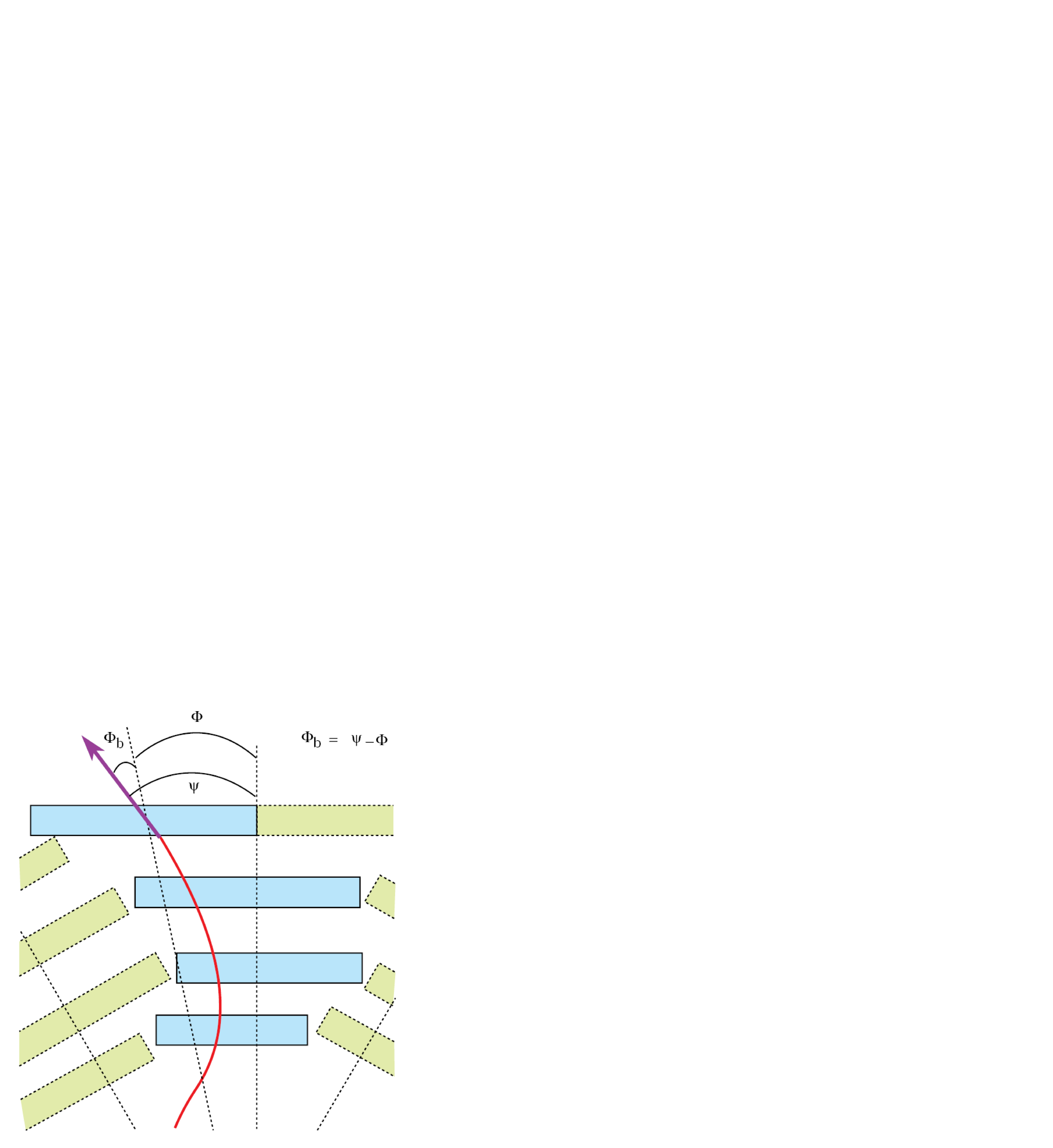}
    \caption{Definition of the $\Phi$ and $\Phi{\rm _b}$ parameters provided by the DT Local Trigger, shown
for the outermost muon chamber.
The angle $\psi$ represents the inclination of the trigger segment with respect 
to the direction perpendicular to the chamber plane. The angle $\Phi$ is defined with respect to the center of the chamber.}
    \label{fig:figure3}
  \end{center}
\end{figure}

The position and angle parameters of the trigger segment, named $\Phi$ and $\Phi{\rm _b}$ respectively, are 
illustrated in Fig.~\ref{fig:figure3}, together with the angle $\psi$ that
defines the inclination of the trigger segment with respect to the direction perpendicular
to the chamber plane.
The parameter $\Phi$ is determined by the position of the 
trigger segment with respect to the center of the chamber, whereas 
the quantity $\Phi{\rm _b}$ is the angle between the direction of the trigger segment and the
straight line from the interaction
point to the segment position. 

The data from the four Trigger Server units of every muon sector are sent to the Regional Trigger~\cite{ref_dttf}, 
which independently performs the matching between trigger 
segments in
the $\phi$ and $\theta$ views, using two different systems: the Phi Track Finder and the Eta Track Finder,
respectively. 
In the data analysed for this study the Eta Track Finder was not yet operational. The matching of segments between chambers 
is performed by the Phi Track Finder using configurable look-up tables, 
which define the range, in terms of position and angle, 
where a valid extrapolation of a trigger segment from one chamber to another is expected.
The Phi Track Finder extrapolation mechanism was not used during the data taking with 
cosmic ray muons in order to maximize the geometrical acceptance.
The Phi Track Finder processed data within one sector or two neighboring sectors, by searching 
for a signal coincidence in at least two chambers. Up to four candidates, 
ordered according to their transverse momenta,
are selected and forwarded to the Global Muon Trigger, 
where they are combined with muon-trigger candidates from RPCs and CSCs~\cite{trig_TDR}.
During data taking with cosmic rays,
a Level-1 trigger rate of about 240 Hz was passed from the Phi Track Finder to the Global Muon Trigger.

The DT Local Trigger signals from the various chambers must arrive synchronously at the Regional Trigger,
and a dedicated synchronization procedure was developed for triggering with cosmic rays.
The time taken by the particles traversing the detector from top to bottom depends on the track incident angle, 
and is approximately 50 ns for cosmic muons traversing both sectors 4 and 10.
To synchronize the DT trigger system, the latency of the trigger signals of the chambers in the upper sectors 
was increased up to a maximum delay time
of 50 ns, which corresponds to two bunch crossings. 
The delays for each individual chamber were obtained with data from dedicated synchronization runs by examining 
the bunch crossing distribution of the trigger segments from different detector regions with respect to a 
reference sector. This allowed the synchronization of all chambers to be 
within a fraction of a bunch crossing, 
regardless of the 
direction and the position of the incoming particle.

\subsection{The Event Selection}
Unless explicitly mentioned, the results shown in this paper are obtained using 
data collected with the CMS solenoid operating at $\rm 3.8\, T$. 
The data were collected in periods of stable DT operation.
No requirements on the status of the other CMS detectors and their participation in the data taking 
are applied to the event selection, except when explicitly mentioned. 
The sample used for the analysis 
corresponds to approximately 10 million events. This is large enough to make statistical uncertainties negligible
for the measurements performed.

The trigger and data acquisition are synchronized with a 40 MHz 
signal replicating the LHC BX frequency.
The Level-1 Trigger signal requires the coincidence, at the same BX, of at least two trigger primitives 
from two chambers of the same sector 
or of two adjacent sectors.
To avoid any bias in the event selection, it is required that at least two chambers distinct from the one under investigation have provided
trigger primitives. This guarantees that the event is selected by the Level-1 Trigger 
independently of the chamber under investigation.
To measure the trigger efficiency in a chamber, the presence of a
local track segment reconstructed with at least 7 out of 8 TDC hits
in the $\phi$ view is also required. The radial angle of the track segment must be in the range $| \psi | < 30^\circ$, 
to stay within the TRACO angular acceptance. 
To remove events with two or more muon tracks in the same chamber, 
at most 12 TDC hits are allowed in the $\phi$ view of the chamber 
under investigation.
The efficiency of the DT Local Trigger in the chamber is defined as the fraction of the selected events
that have an associated trigger primitive in the chamber, in the same or in any of the two adjacent BXs.
This definition of efficiency
allows the measurement of the effective net trigger capability, excluding the geometrical 
acceptance and the DT cell efficiency. 
This event selection is employed for the analysis described in this paper, 
except where explicitly stated otherwise.

\section{Performance of the DT Local Trigger}
\label{performance}

Muons produced in LHC collisions will cross each DT chamber at a well defined time with respect 
to the signal of the BX clock, with a jitter of a few nanoseconds, 
due to the different path lengths in the magnetic field. The 
BTI electronics will be properly synchronized with the arrival time of the particles 
to perform the best hit alignment~\cite{sync, CRAFT_sync}.
However, cosmic ray muons traverse the detector at arbitrary times 
with respect to the 40 MHz signal that simulates the BX clock.
Taking the signal of the BX clock as a time reference,
the BTI is configured to detect hit alignments in the most efficient way
in a time range that is about $10\,\rm{ns}$ wide.
Therefore, only muons arriving within the $\rm 10 \,ns$
time window will be correctly identified by the trigger, while
outside this time window the efficiency of the BTI to find aligned hits is degraded.

\subsection{Trigger Efficiency}

\begin{figure}[hbtp]
  \begin{center}
    \includegraphics[width=7cm,angle=90]{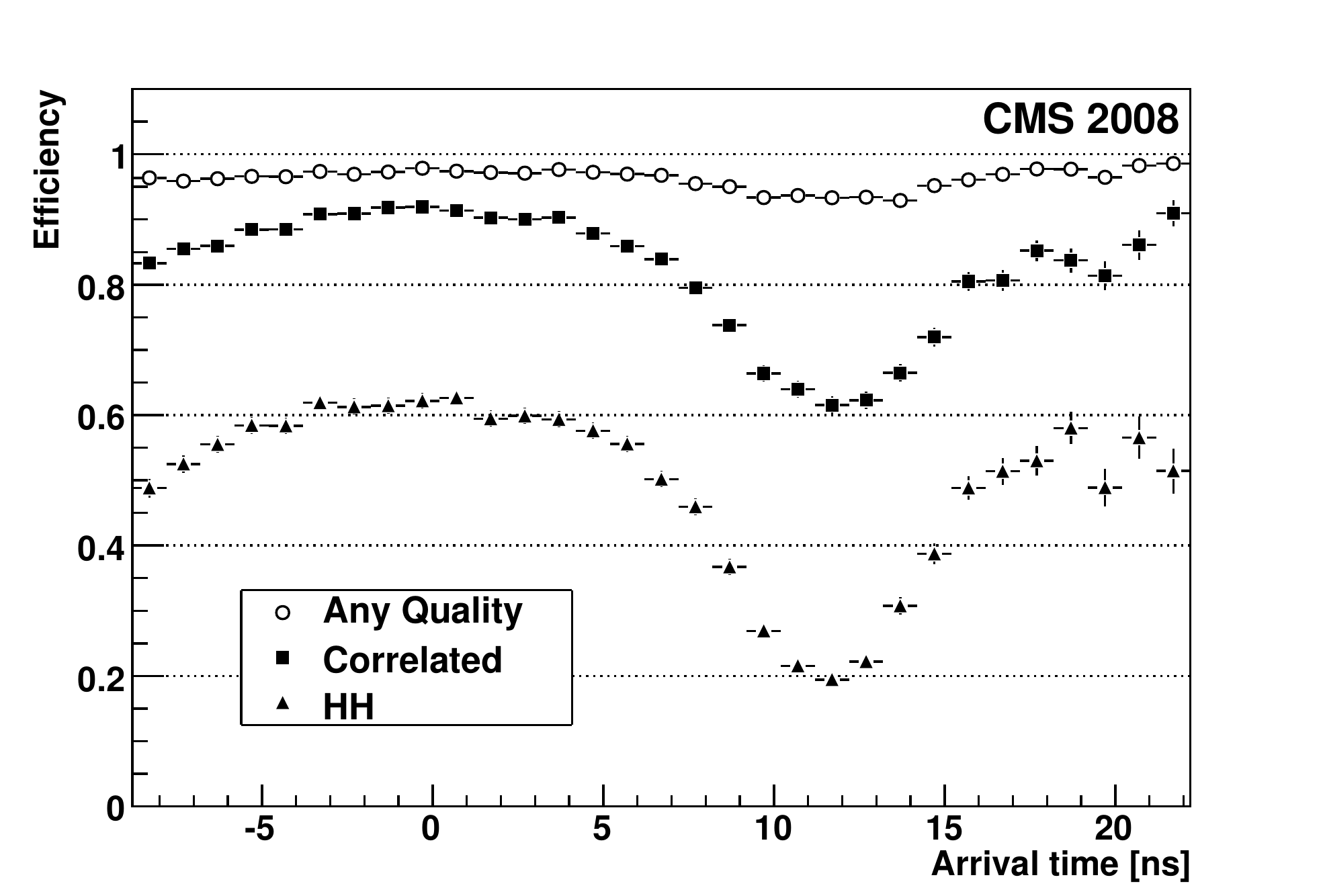}
    \caption{Efficiency of the DT Local Trigger as a function of the muon arrival time 
in the chamber, defined with respect to the 
$t_{\rm Trig}$ time reference. Results for a
chamber of type MB1 are shown as an example, for various types of trigger quality. }
    \label{fig:figure4}
 \end{center}
\end{figure}

To illustrate the effect of the random arrival time of the cosmic rays on the performance measurements presented in this paper,
Fig.~\ref{fig:figure4} shows the DT Local Trigger efficiency 
as a function of the arrival 
time of the muon with respect to the global time reference $t_{\rm Trig}$.
The quantity $t_{\rm Trig}$ represents the value to be subtracted from the  
time measured by the TDC (called $t_{\rm Digi}$) to correct for the
trigger latency and for 
differences in the propagation time of the signals in the system.
The determination of $t_{\rm Trig}$ is described in Ref.~\cite{calibration}, and 
it is performed by fitting the position of the rising edge of the $t_{\rm Digi}$ distribution separately in each superlayer. 
The arrival time with respect to $t_{\rm Trig}$ 
is individually computed for every 
muon and event, by performing
a linear fit to the TDC hits in the chamber. The arrival time 
is a free parameter of the fit, and it is 
determined, event-by-event, as the time shift 
to be added to $t_{\rm Trig}$ giving the 
best spatial resolution for every event by
achieving the best 
track-segment fit~\cite{tzeroev}. 
This shift represents the time offset at which the muon traversed the chamber with 
respect to the reference time $t_{\rm Trig}$, which is the mean arrival time averaged over many events.

%For each muon, the arrival time with respect to $t_{\rm Trig}$ is computed event-by-event
%from a linear fit to the TDC hits in the chamber, and it is 
%the offset to be added to $t_{\rm Trig}$
%to achieve the best segment fit~\cite{tzeroev}.

Figure~\ref{fig:figure4} shows that the DT 
Local Trigger efficiency is not constant 
as a function of the arrival time of the particle, 
in particular for trigger segments of HH quality, 
but it is maximal and flat in a region approximately $10 \,\rm{ns}$ wide. 
To first order, the efficiency has a periodicity of $25\,\rm{ns}$, which is the time difference 
between two consecutive BXs. As the events selection requires the time 
coincidence of trigger segments in at least two chambers,
uncertainties in their relative synchronization,  which was typically of the order of a fraction of 
a BX during the period of data taking, can bias the measured arrival time of the muons, thus resulting in an 
imperfect periodicity of the efficiency curve. The time dependence of the efficiency reflects the fact that the trigger system is optimized 
for muons arriving in the time range corresponding to the 
flat-top region of Fig.~\ref{fig:figure4}. For LHC operation, the trigger timing will be adjusted
so that 
muons from collisions will traverse the detector within this time range.
For the tests with cosmic ray muons, however, 
this results in an overall degradation of the average performance of the DT Local Trigger 
compared to the expectations for the
LHC and the results with bunched-beam tests, in terms of efficiency, quality of the
trigger primitives, and BX identification.
 
In the $10\,\rm {ns}$ flat-top region, where the trigger system 
is well synchronized with the incoming particles, 
the DT Local Trigger efficiency is approximately 0.60 for triggers of HH quality, 
0.90 for correlated triggers, 
and 0.97 for any trigger quality.
These values are consistent with previous measurements at bunched-beam tests~\cite{NIM_2004, NIM_2007},
as well as with the technical specifications required for efficient triggering with particle collisions~\cite{trig_TDR}.

The average DT Local Trigger efficiencies obtained using all cosmic
ray muons, regardless of their arrival time with respect to the reference
time $t_{\rm Trig}$, are 0.95 for triggers of any quality, 0.80 for
correlated triggers, and 0.48 for triggers of HH quality. These values are
lower than the values reached in the plateau of maximum efficiency
shown in Fig.~\ref{fig:figure4}, where the performance of the system is optimal. They
have a spread of 0.02 among the chambers, mostly caused by the
different angles of incidence of the cosmic rays on the chambers and by
a few malfunctioning trigger devices. The few chambers with major
hardware or read-out problems are not taken into account in the
computation. These problems were fixed during the detector shutdown
at the end of 2008.

\begin{figure}[hbtp]
  \begin{center}
    \includegraphics[width=7cm, angle=90]{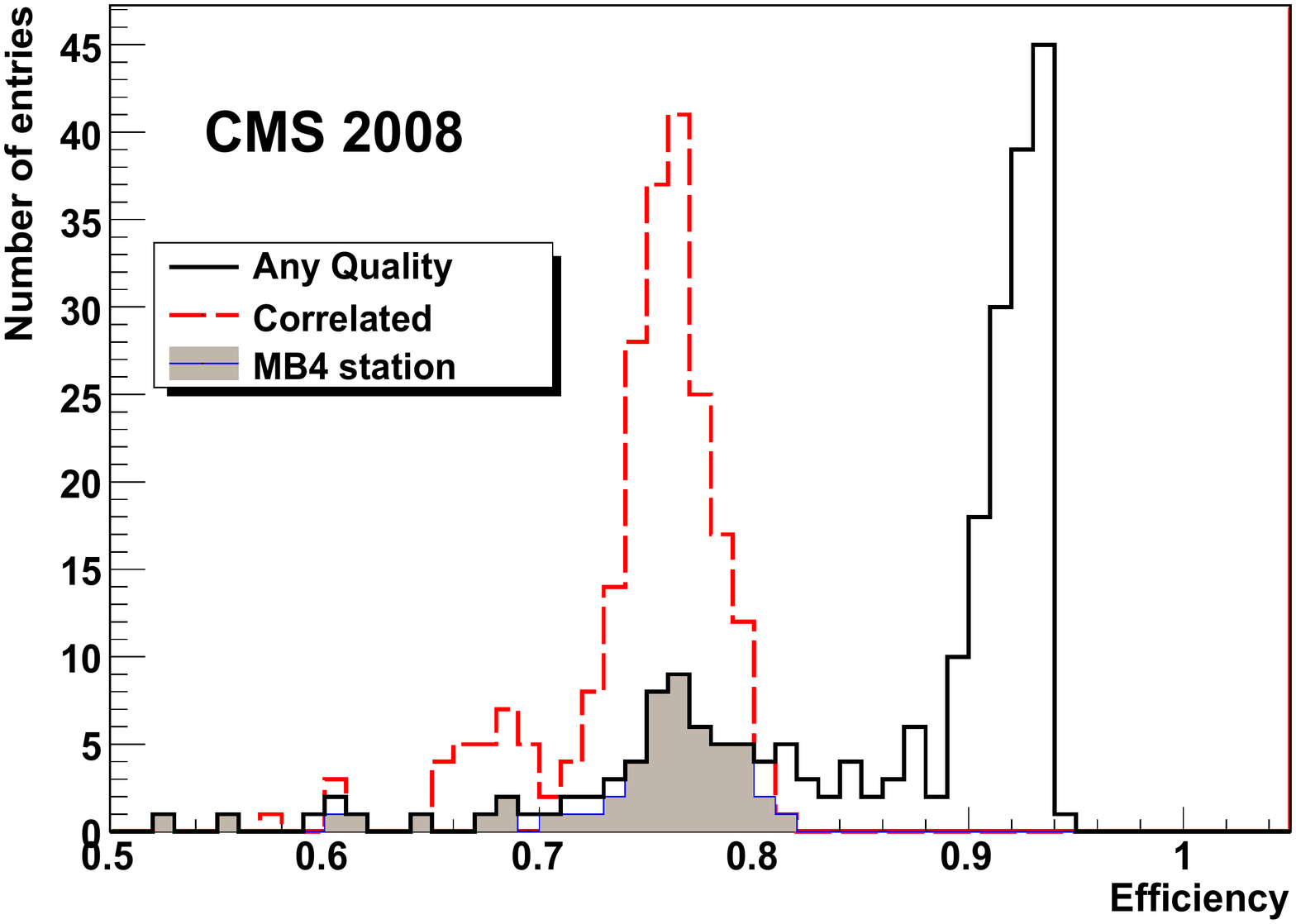}
    \caption{Distribution of the DT Local Trigger efficiency in the $\phi$ view of the muon chambers,
for events selected by the RPC trigger.
Each entry corresponds to one barrel chamber.
Results for any trigger quality and for correlated triggers 
are shown. The shaded histogram represents the 
results for station type MB4, for which only correlated triggers are provided.}
    \label{fig:figure5}
  \end{center}
\end{figure}

The results on the DT Local Trigger efficiency are cross-checked using an independent event selection that does 
not explicitly require the presence of DT hits in the chamber under study. 
For this purpose, events triggered by the 
RPCs are selected. In such events, muon tracks are subsequently reconstructed using the DT hits in the muon barrel and CSC hits 
in the endcaps. The tracks are extrapolated along the 
detector, and required to cross the chamber under study in its active region, without any
requirement on the 
presence of DT hits in that chamber. The chamber is considered efficient if there 
is a trigger segment in the $\phi$ view, in the same or in any of the two adjacent BXs.
In order to remain within the 
geometrical acceptance of the trigger, only tracks with an angle of incidence $| \psi | < 30^\circ$ were considered.
Figure~\ref{fig:figure5} shows the distribution of the DT Local Trigger 
efficiency measured for these events. Each entry 
represents one muon chamber. Results are shown for any trigger quality
and for correlated triggers. Results for station type MB4, for which only correlated triggers are 
provided, are shown separately. In contrast to the previously described measurement, 
this definition of the DT Local Trigger efficiency includes effects of geometrical 
acceptance at the edges of the chambers, cell inefficiency, dead channels, and
hardware failures. Therefore, the measured values are 
on average a few percent lower than the ones previously reported. 
For triggers of any quality the efficiency distribution has a maximum at $\rm 95 \,\%$.
Values below $\rm 70 \,\%$ are due to read-out or hardware problems, which have been fixed 
during the detector shutdown at the end of 2008.
For correlated triggers, the efficiencies below $\rm 70 \,\%$
are related to chambers oriented vertically in the detector (sectors 1 and 7).  In this case, given the
angular distribution of cosmic rays, there is a high probability
to fail the correlation between superlayers near the edges of the chambers. 
In this detector region the low efficiency is partially recovered if trigger segments of any quality are accepted,
and in this case it reaches values between $\rm 80 \,\%$ and $\rm 90 \,\%$.

\begin{figure}[hbtp]
  \begin{center}
    \includegraphics[width=6cm, angle=90]{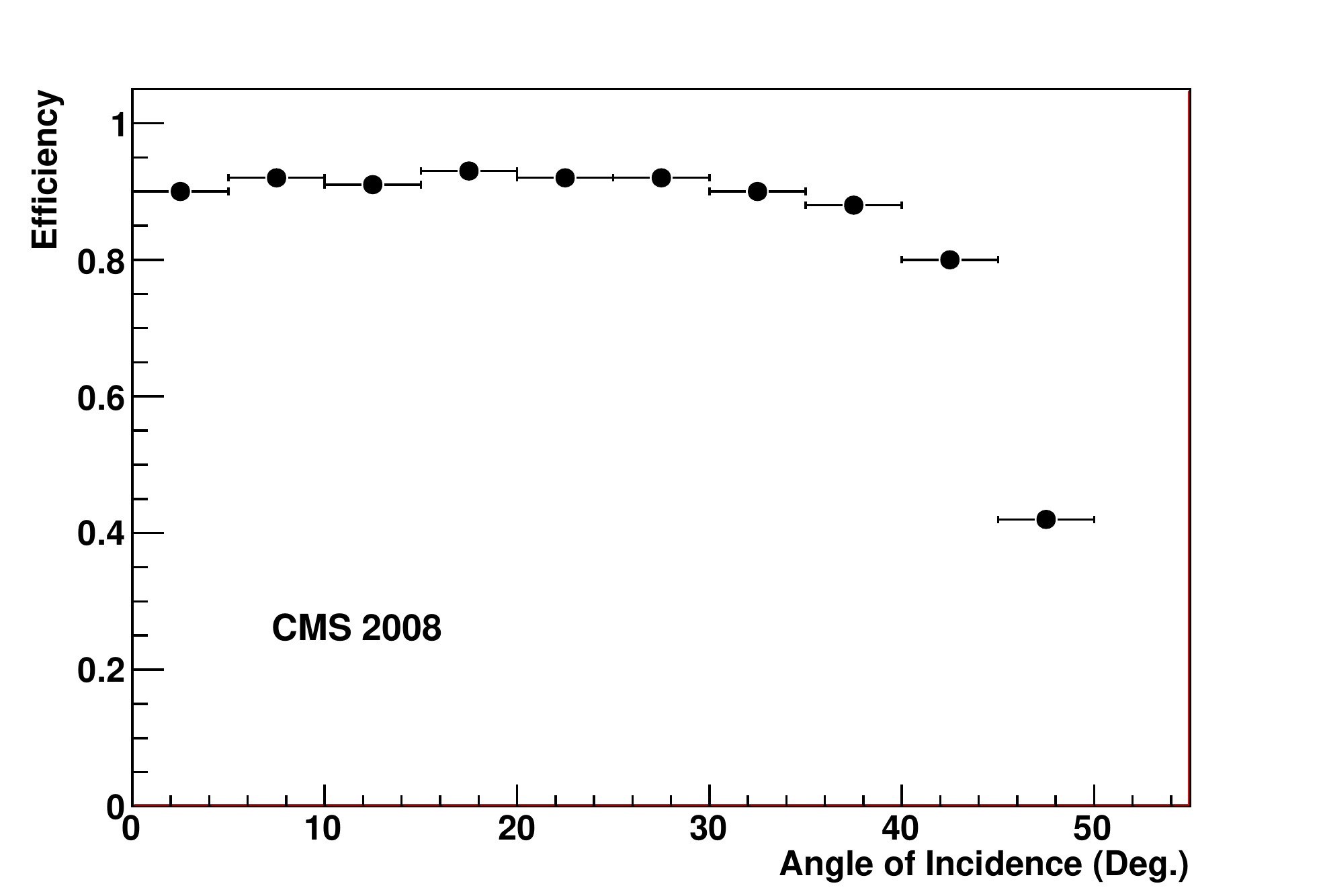}
    \caption{Average DT Local Trigger efficiency in the $\phi$ view
of the muon chambers for trigger segments of any quality, as
a function of the angle of incidence $| \psi |$ of the muon tracks.}
    \label{fig:figure5_2}
  \end{center}
\end{figure}

The DT Local Trigger efficiency as a function of the incidence angle $|\psi|$ is also
measured, for trigger segments of any quality.
In this case the requirement on the incidence angle of the track, $| \psi | < 30^\circ$, is removed,
and the efficiency value is averaged over all chambers, with the exclusion of sectors 1 and 7.
The results, reported in Fig.~\ref{fig:figure5_2},
are in agreement with previous test beam measurements~\cite{NIM_2004} 
and show that the trigger efficiency
has a constant plateau for tracks with inclination up to about $\rm 35^\circ$, and 
it decreases by a factor of 2 at the maximum BTI angular acceptance of
$\rm 45^\circ$.
This corresponds, in the case of LHC collisions, to full trigger efficiency 
for tracks with transverse momenta exceeding 3 GeV/$c$.

\subsection{Bunch Crossing Identification}

\begin{figure}[hbtp]
  \begin{center}
    \includegraphics[width=5.2cm,angle=90]{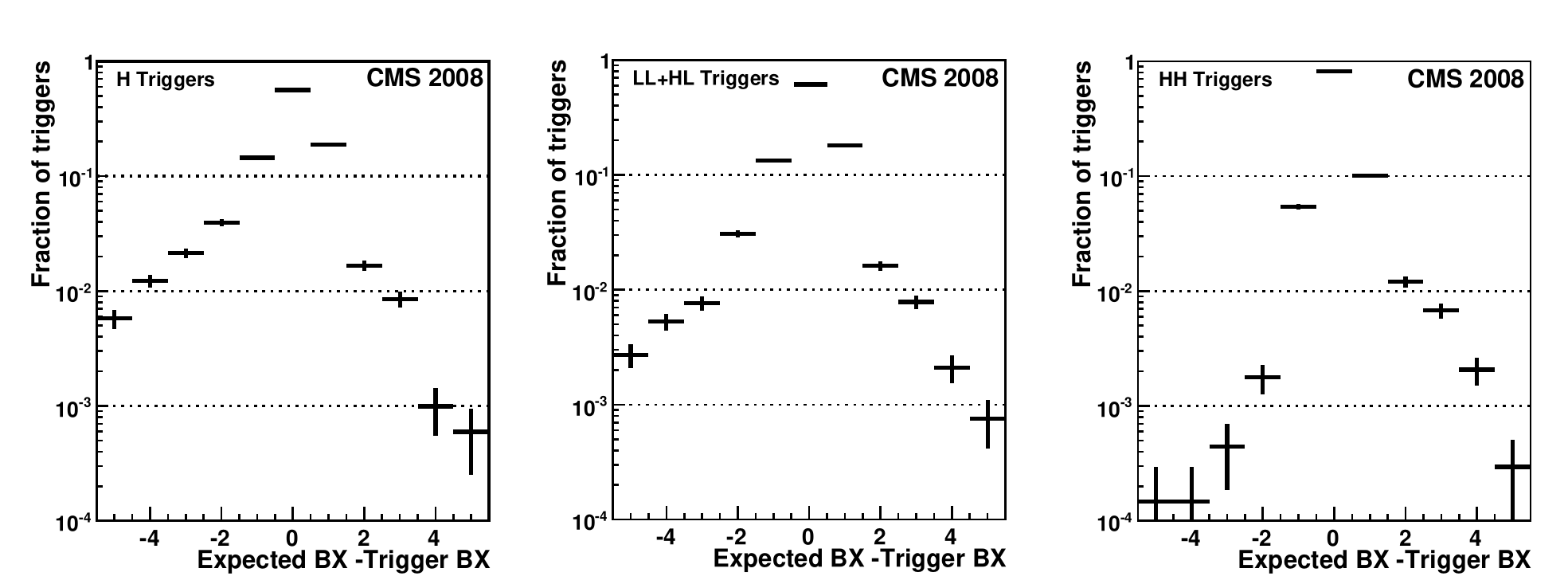}
    \caption{Distribution of the difference between the ``expected BX'' and the observed BX 
( ``trigger BX'') of the trigger segment, in one chamber of type MB1 in sector 4, 
for different trigger qualities. The distributions are normalized to unity. }
    \label{fig:figure6}
  \end{center}
\end{figure}

The performance of the BX identification is measured by comparing the assigned bunch 
crossing in the chamber under study, called ``trigger BX'', with respect to the BX defined by
the coincidence of the other two
trigger segments in the same sector, called  ``expected BX''.
The difference between the ``expected BX'' and the ``trigger BX'' is shown in Fig.~\ref{fig:figure6} for different trigger qualities.
The data from a chamber of type MB1 in sector 4 are used as an example. The results are similar for 
the other chambers in the detector. The events for which the difference between the ``expected BX'' and the ``trigger BX'' is zero
have a correct BX identification.
The distributions, normalized to unity, show that, even in the difficult environment 
of cosmic rays, triggers of HH quality provide correct 
BX identification in $\rm 82 \,\%$ of the cases. The distribution is affected by imperfect synchronization between the chambers during data taking.
The distribution is wider for poorer trigger quality. Because of a feature of the BTI system, the events with badly identified BX mainly populate the left-most part of the 
distribution for low-quality triggers, 
as they are trigger segments with wrong hit alignments that are typically produced later with respect to the correct BX. 
Such triggers are significantly suppressed in the case of quality type HH. 

\begin{figure}[hbtp]
  \begin{center}
    \includegraphics[width=7cm,angle=90]{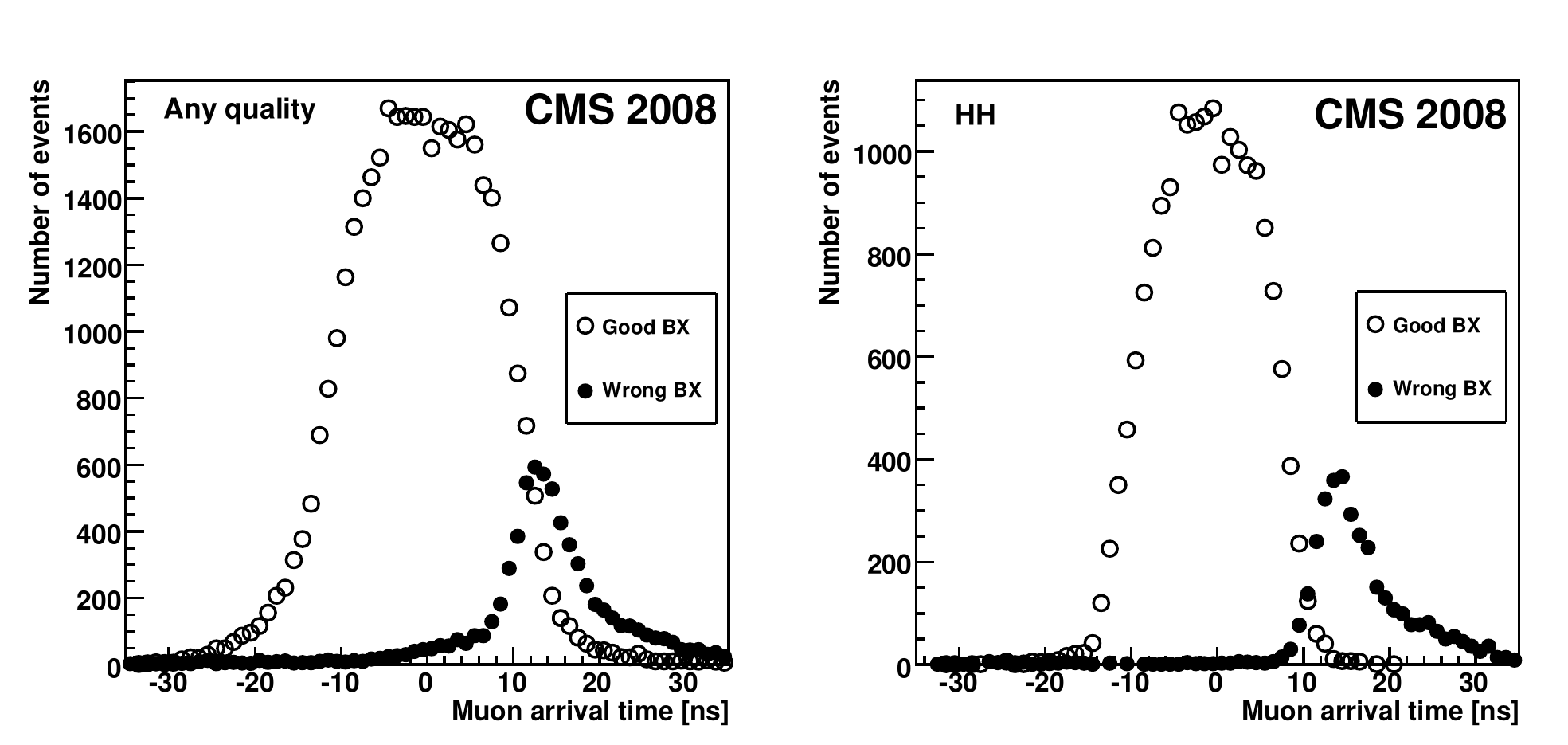}
    \caption{Left: distribution of the muon arrival time, for trigger segments of any quality.
Right: distribution of the muon arrival time for trigger segments of HH quality. 
Events with correct and wrong BX identification are shown separately.  Results for a chamber of type MB1 in sector 4 are shown as an example.
}
    \label{fig:figure7}
  \end{center}
\end{figure}

In bunched-beam tests the BX identification for trigger segments of quality 
HH was close to  $\rm 100 \,\%$~\cite{NIM_2004, NIM_2007}.
In the present analysis, the degraded performance of the BX identification is explained by the
timing characteristics of the cosmic rays already discussed.
Events selected with the optimal arrival time of the muon tracks 
(as discussed in section 3.1) show that
the BX identification capability for triggers of HH quality approaches $\rm 100 \,\%$, thus 
matching the design expectations for proton-proton collisions.
This can be seen in Fig.~\ref{fig:figure7}, which shows the distribution
of the arrival time of the muon tracks (in the same chamber used for Fig.~\ref{fig:figure4}), 
for events with a trigger segment of any quality (left) and for quality type HH (right). 
The distributions for trigger segments with correct BX assignment (``expected BX'' equal to ``trigger BX'')
and wrong BX assignment are shown separately. In the region 
approximately $\pm\,5 \,\rm{ns}$ wide around zero, 
where the timing of the trigger system is optimized (as
discussed for the results of Fig.~\ref{fig:figure4}), the wrongly timed triggers of any
quality contribute only marginally, and they can be completely neglected
in the case of trigger segments of HH quality. 

\subsection{False Dimuon Trigger Rate}

The DT Local Trigger provides up to two trigger segments for any BX in the $\phi$ view of each chamber.
This feature was introduced to 
maximize the trigger 
efficiency for nearby muons. On the other hand, a ghost suppression mechanism 
is implemented to reject false trigger 
pairs that originate from a single muon crossing the detector. As the dimuon production rate 
above any given transverse momentum threshold is expected to be approximately $\rm 1 \,\%$ of the
single muon rate at the LHC~\cite{trig_TDR},
the rate of false Level-1 dimuon triggers must be well below $\rm 1 \,\%$.
With cosmic rays the fraction of double triggers over single triggers provided in a single chamber at 
the same BX, averaged over all the chambers of type MB1, MB2, and MB3, is \mbox{0.03 $\pm$ 0.01}.
The DT Local Trigger in station type MB4
provides a lower 
fraction of double triggers, namely \mbox{0.012 $\pm$ 0.005}, since only correlated 
trigger segments are provided by this station type.
The uncertainty reflects the observed spread of values among the various chambers, due to systematic effects 
like noisy channels, or different angles of incidence of the cosmic rays.
The results are consistent with  
test beam measurements~\cite{NIM_2004, NIM_2007}. The false dimuon Level-1 trigger rate in the barrel region is further
suppressed, as the matching of at least two segments in two different
chambers is required to form a trigger candidate, and as a further selection
is applied by the Global Muon Trigger. The measured false dimuon trigger rate per
chamber is significantly smaller compared to the assumptions made in
Ref.~\cite{trig_TDR}. Hence, the false dimuon Level-1 rate in the barrel region is
estimated to be below the 0.1 \% reported in Ref.~\cite{trig_TDR}, which is at least one order of 
magnitude smaller than the real dimuon rate and can, therefore, be
neglected.

\begin{figure}[hbtp]
  \begin{center}
    \includegraphics[width=7cm, angle=90]{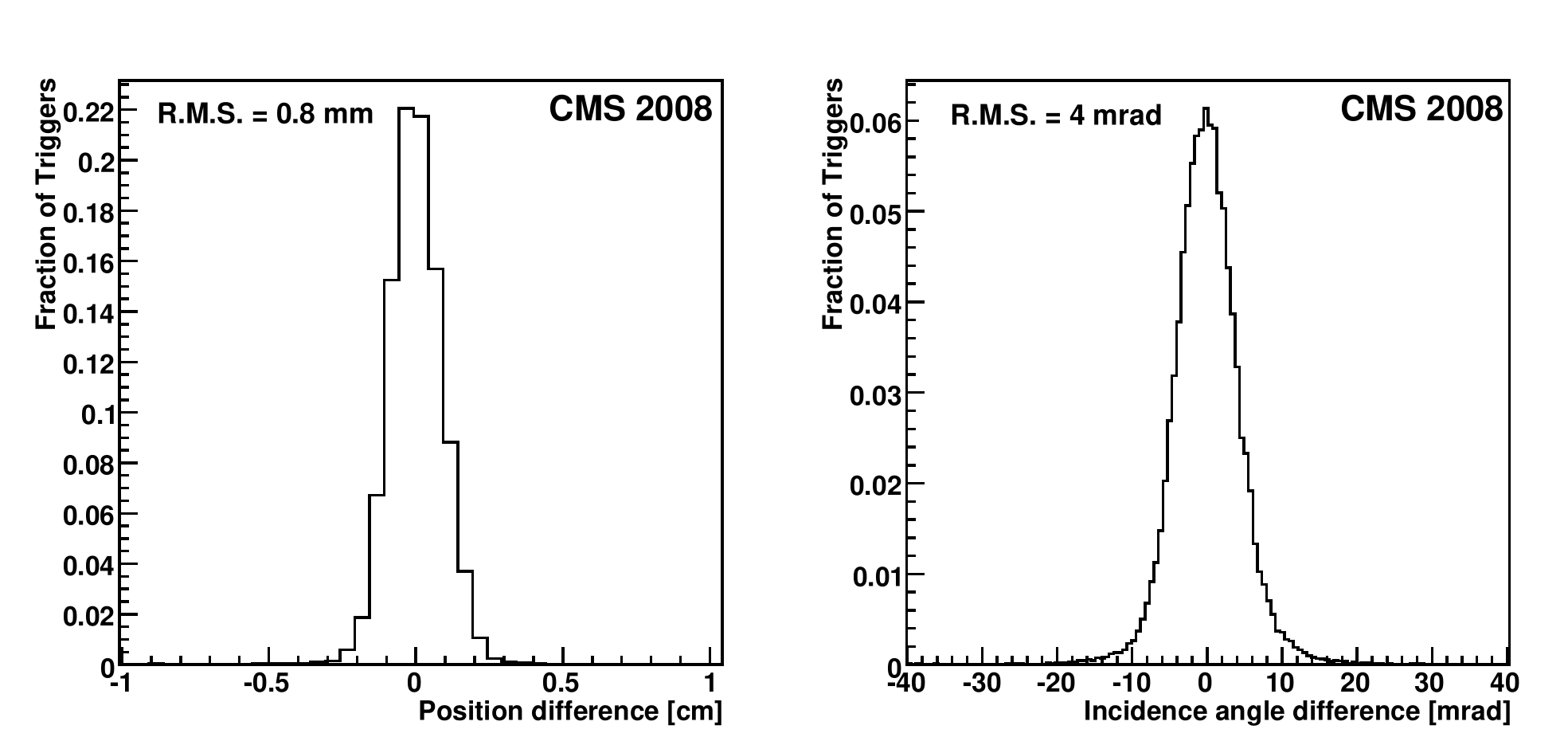}
    \caption{Left: distribution of the difference between the position of the local track segment and the trigger 
segment.
Right: distribution of the difference between the incidence angle measured by the local track segment and the 
trigger segment. Results are shown for a chamber of type MB1.}
    \label{fig:figure8}
  \end{center}
\end{figure}

\subsection{Position and Angular Resolution}

The track segments obtained by fitting the TDC information in each chamber 
and used for \mbox{offline} muon reconstruction
provide an accurate determination of the position and the incidence angle of the muon in the chamber 
independent of the trigger output.
These quantities are compared with the corresponding information assigned by the DT Local Trigger 
to the trigger segments, $\Phi$ and $\Phi{\rm _b}$. Figure~\ref{fig:figure8} (left) shows the distribution of 
the difference between the position computed by the reconstructed track segment and by the trigger segment for 
a chamber of type MB1 taken as an example.
The RMS of the distribution is $\rm 0.8 \, mm$. This provides
an estimate of the position resolution of the trigger segment, 
which is the same for every station type.  This measurement is 
in agreement with previous test beam results~\cite{NIM_2004}. 
The uncertainty related to the track position measurement is neglected.

A similar study is performed on the track incidence angle $\psi$. 
Figure~\ref{fig:figure8} (right) 
shows the distribution of the difference 
between the incidence angle of the reconstructed track and 
the trigger segment.  The RMS of the distribution is 4 mrad. Taking into account the 
slight reduction of the angular resolution arising from the 
timing characteristics of the cosmic rays, the resolution is 
in agreement with the value of 3 mrad obtained in previous test beam measurements~\cite{NIM_2004}. 
The result guarantees that
the expected performance in terms
of position and transverse momentum resolution at the output of the 
Level-1 trigger will be achieved for collision data taking~\cite{trig_TDR}.

\subsection{Effect of the Magnetic Field}

In station type MB1 of the 
wheels 
$+$2 and $-$2, the magnetic field is expected to affect the drift-path lengths 
such that the 
maximum drift time is increased by $\mathcal{O}$(10 ns)~\cite{Bfield}.
A variation of the maximum drift time, which corresponds to an apparent change of the drift velocity,
could degrade the DT Local Trigger performance, affecting in particular the
BX identification. A sample of approximately 5 million events collected with no magnetic field is
used to study this effect. The means of the 
BX distributions determined by the 
DT Local Trigger in a chamber (obtained as in Fig.~\ref{fig:figure6}) are compared for data taken 
with and without the
magnetic field.

Some corrections are applied to compare the results obtained from the two data sets.
As muons are bent by the magnetic field,
the timing of the trigger signals between the various chambers and sectors 
is slightly affected because of
the different path length, especially for low momentum particles.
In addition, some further adjustments in the synchronization between the various
sectors were performed between the two data taking periods. Therefore,
a direct comparison of the means of the BX distributions obtained in the two conditions is 
biased by these systematic effects. As in station type MB4 the effect of the magnetic field on the drift velocity
is expected to be negligible, it is assumed that in this case
any deviation from zero observed in the difference of the BX distributions is entirely
due to the systematic effects.
Results for station type MB4 are then set to zero, separately in each sector, and the results for the other stations
are shifted by the same amount to correct for the systematic effects.

\begin{figure}[hbtp]
\begin{center}
  \includegraphics[width=7cm,angle=90]{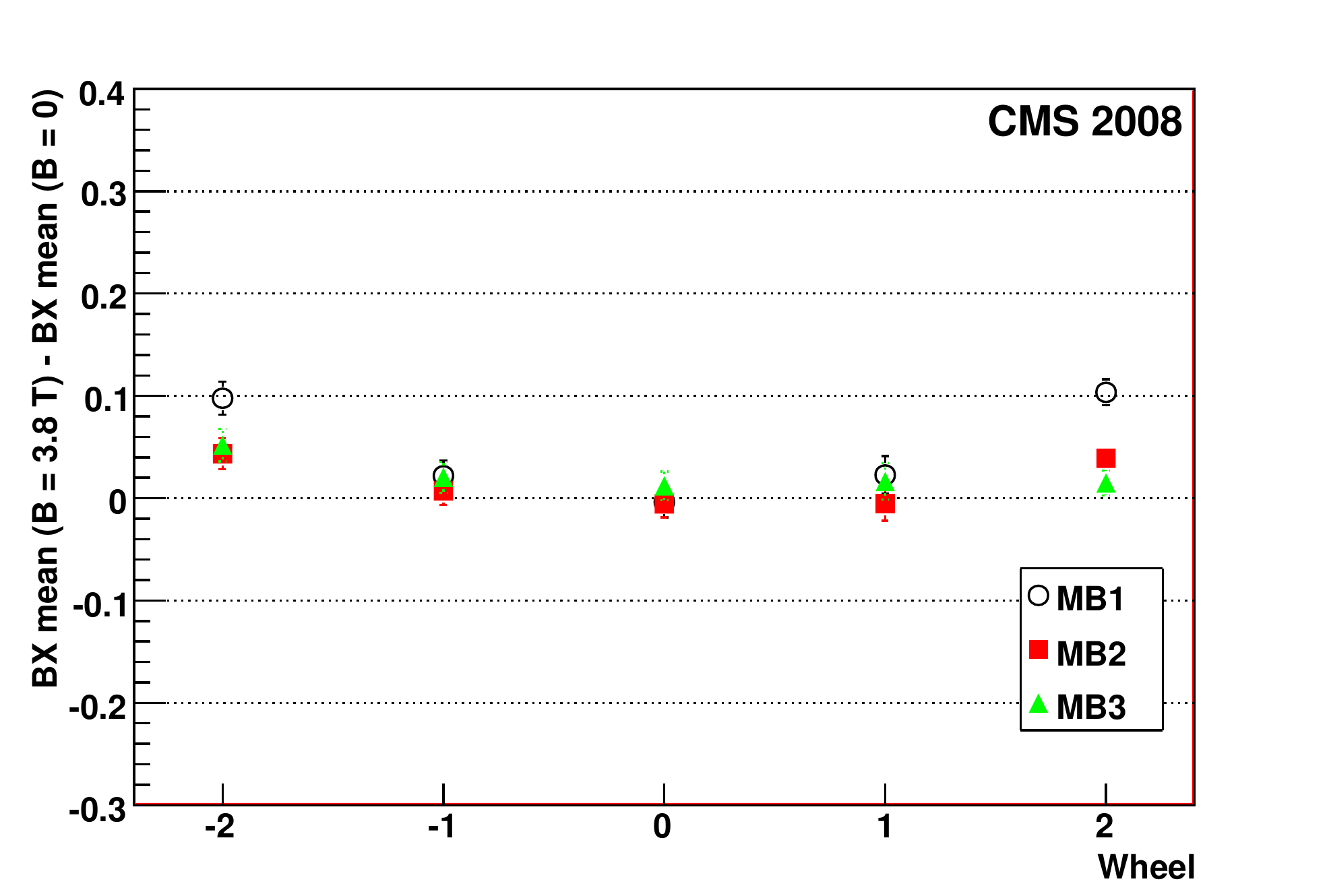}
  \caption{Difference between the means of the BX distributions determined by the DT Local Trigger 
with and without magnetic field, in units of BX,
    as a function of the wheel number, for MB1, MB2 and MB3 station types. }
  \label{fig:figure9}
\end{center}
\end{figure}

Figure~\ref{fig:figure9} shows the difference between the means of the BX distributions, with and without the magnetic field, 
as a function of the wheel number for different station types, after the correction procedure described above.
The results are averaged over the various sectors, and only statistical uncertainties are shown in the figure.
Small deviations from zero are observed in all stations and wheels, and they are partially 
due to the aforementioned systematic uncertainties, 
which could not be completely suppressed. 
The largest deviation from zero is observed in MB1, at the 
wheels $+$2 and $-$2, where the magnetic field is indeed 
expected to have the largest effect on BX identification. 
Comparing results with and without the magnetic 
field, the maximum average BX displacement 
is 0.1 units of a BX, corresponding to about $3 \,\rm{ns}$. 
This shift is much smaller than $25\,\rm{ns}$, which is the time between two consecutive BXs, and thus
it cannot significantly affect the performance of the trigger~\cite{Bfield}.

A study of the BX displacement as a function of the position of 
the trigger candidate along the $z$ coordinate shows that the BX assignment is 
fairly constant across the muon chambers, increasing at the edges 
with the maximum value of $5\, \rm{ns}$, which is sufficiently small compared to $25\,\rm{ns}$.
Consequently it is possible to 
configure the BTIs with the 
same value of the
drift-velocity parameter in each chamber, regardless of
its location in the detector or the magnetic field.

The comparison between events recorded with and without the magnetic field 
is used to study the effect of the field on 
the position and angular resolution of the trigger segments. The results 
show that the position and direction determination by the DT Local 
Trigger is not affected by the presence of the 
magnetic field. 
The comparison of data taken with and without the magnetic field also shows that
the magnetic field does not affect the trigger efficiency, 
except in station MB1 of the wheels $+$2 and $-$2 where the fraction of correlated triggers is decreased by less than $\rm 2 \,\%$.

\section{Comparison with the Emulator}
\label{emulator}
\begin{figure}[hbtp]
  \begin{center}
    \includegraphics[width=7cm,angle=90]{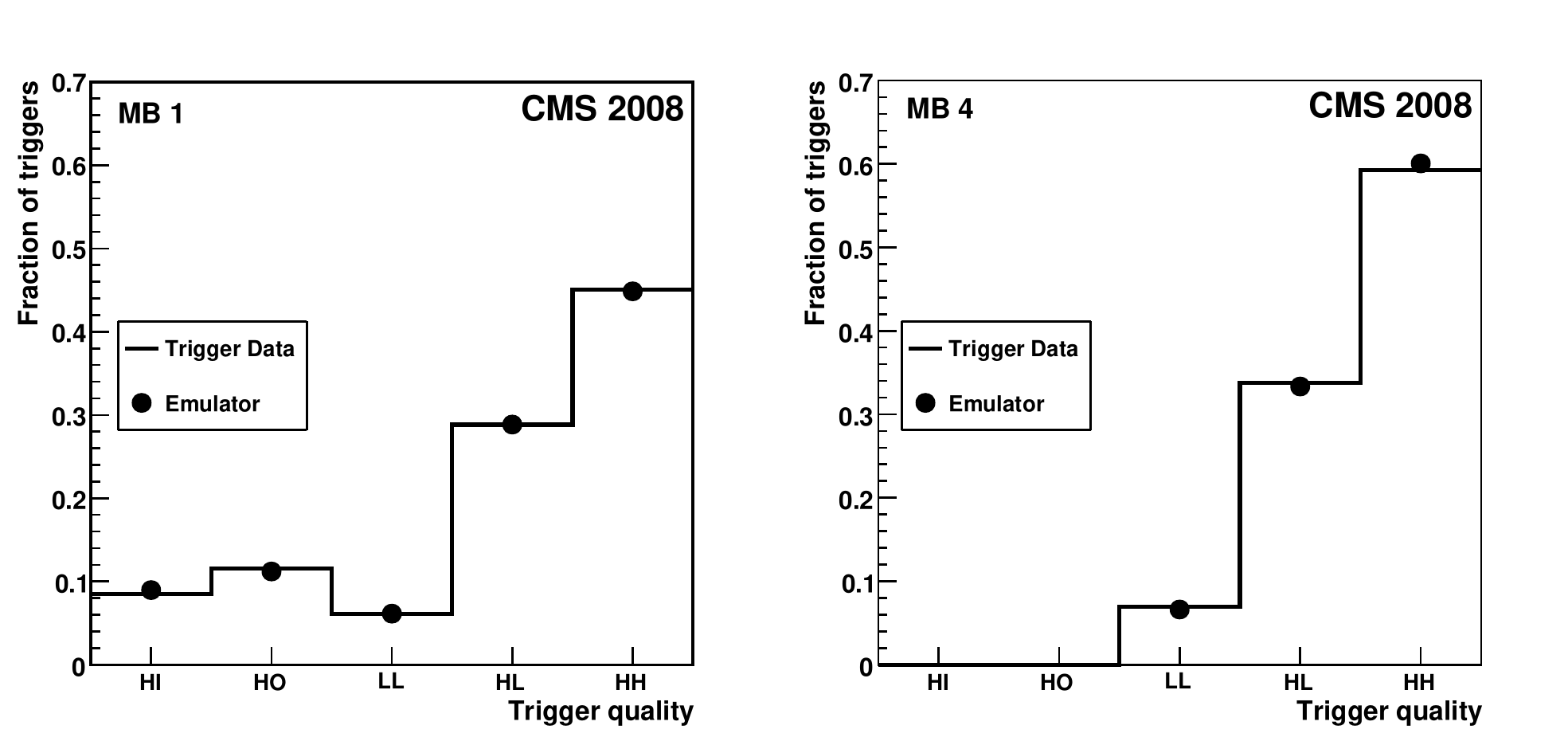}
    \caption{Distribution of the quality of the trigger segments, for data and emulation, 
for station types MB1 (left) and MB4 (right).}
    \label{fig:figure10}
  \end{center}
\end{figure}

A detailed simulation of the DT Local Trigger selection algorithm is implemented in 
the CMS software framework, and is 
part of the emulator program that reproduces the output of the entire Level-1 trigger.
The emulator processes the signals from the drift cells: for each cell 
the difference $t_{\rm Digi} - t_{\rm Trig}$ is computed and sent as input to the emulator.
The quantity 
received 
as input by the trigger electronics is not exactly the same as $t_{\rm Digi} - t_{\rm Trig}$,
although its physical meaning is the same.
In fact, the BTIs 
are directly linked to the wires of the DT chamber, receiving discriminated signals on 
the on-board electronics, while the quantity $t_{\rm Digi}$ and the pedestal parameter $t_{\rm Trig}$
are determined by the TDCs (which have a different hardware
connection to the DT cells with respect to the BTIs) and by an offline analysis, respectively.
This makes the exact data-emulator agreement impossible by design. 
Discrepancies between data and emulation, on an event-by-event comparison, 
are mostly observed in the quality of the trigger primitives.
Even considering this intrinsic limitation, an event-by-event comparison shows that,
among all the trigger primitives provided by the trigger electronics, 
more than $\rm 90 \,\%$ are perfectly reproduced by the emulator in all their characteristics.
If the comparison between data and emulator
is performed on a statistical basis rather than event-by-event, the differences between 
data and emulator largely compensate, and the agreement approaches $\rm 100 \,\%$.
This can be seen in Fig.~\ref{fig:figure10}, which shows the comparison of the 
distribution of the trigger quality for station
types MB1 and MB4.

The emulator program assumes a perfectly working trigger system, so that any 
significant discrepancy with the data is 
an indication either of malfunctioning electronics or mis-configuration.
On the other hand, dead cells, high-voltage problems, noisy cells, or any other malfunctions
affecting the
presence of the digitized information from the TDCs, but
not related to the trigger devices, are
always reproduced by the emulation, resulting in a very good matching with the data.
Therefore, the emulator is a very powerful tool for use during data taking for 
diagnosing problems that are related to the trigger system.

\section{Conclusions}
\label{conclusions}
The muon Local Trigger based on the drift-tube detector of the CMS experiment has been extensively tested 
during the commissioning of the apparatus using cosmic rays. 
Cosmic rays traverse the detector at arbitrary times and with different directions compared 
to particles produced in LHC collisions, for which the muon trigger was designed. 
Therefore, the system performance is degraded with respect 
to the expectations for the LHC, or if compared with measurements previously
performed with bunched beams.
Nevertheless, the device 
operated in a very efficient and reliable way.
An overall efficiency of $\rm 95 \,\%$ is obtained, and $\rm 48 \,\%$ of the trigger segments are of
the best possible quality. 
An accurate measurement of the trigger efficiency as a function of the 
timing of the system shows that, in the time 
window where the trigger electronics is properly synchronized with the arriving muon, 
as it will be the case 
for particles produced at LHC collisions, the DT Local trigger efficiency 
is $\rm 97 \,\%$ for any trigger quality. 
The performance of the BX identification reaches $\rm 100 \,\%$ 
when a correction for the arbitrary arrival time of cosmic ray muons 
is performed.
The rate of false double triggers for single muons is about \mbox{1.5--3.0 $\%$} 
of the single muon triggers in one chamber. 
As the coincidence of trigger segments in at least two chambers is needed by the Phi Track Finder to reconstruct a trigger 
candidate, and as a further selection is applied by the Regional and the Global Muon Trigger,
the false dimuon trigger rate is expected to be at least one order of magnitude smaller than the real dimuon
rate expected from LHC collisions at any transverse momentum threshold.
The position and angular resolution of the trigger segments is $0.8 \, \rm{mm}$ and $4 \, \rm{mrad}$, respectively. 
This performance allows the trigger system to achieve the 
position and transverse momentum resolution needed for efficient operation with collisions.
The magnetic field has a negligible impact on the bunch crossing identification, on the position 
and angular resolution, and on the trigger efficiency,
even in the detector regions where the effect should be largest.
Trigger data were also successfully cross-checked with the emulator.
Although a perfect event-by-event agreement with the data is impossible by design, 
more than $\rm 90 \,\%$ of the trigger primitives are perfectly reproduced by the emulator in all their characteristics.
This makes the emulator a powerful tool for diagnosing problems related to the trigger system during data taking.

\

{\Large
\textbf{Acknowledgements}}

We thank the technical and administrative staff at CERN and other 
CMS Institutes, and acknowledge support from:
FMSR (Austria); 
FNRS and FWO (Belgium); 
CNPq, CAPES, FAPERJ and FAPESP (Brazil); 
MES (Bulgaria); 
CERN; 
CAS, MoST and NSFC (China);
COLCIENCIAS (Colombia); 
MSES (Croatia); 
RPF (Cyprus); 
Academy of Sciences and NICPB (Estonia);
ME, HIP and Academy of Finland (Finland); 
CEA and CNRS/IN2P3 (France); 
BMBF and DESY (Germany); 
BMBF, DFG and HGF (Germany); 
GSRT (Greece); 
OTKA and NKTH (Hungary); 
DAE and DST (India); 
IPM (Iran); 
SFI (Ireland); 
INFN (Italy); 
NRF (Korea); 
LAS (Lithuania);
CINVESTAV, CONACYT, SEP and UASLP-FAI (Mexico); 
PAEC (Pakistan); 
SCSR (Poland); 
FCT (Portugal); 
JINR (Armenia, Belarus, Georgia, Ukraine, Uzbekistan);
MST and MAE (Russia);
MSTDS (Serbia);
MCINN and CPAN (Spain); 
Swiss Funding Agencies (Switzerland);
NSC (Taipei); 
TUBITAK and TAEK (Turkey); 
STFC (United Kingdom); 
DOE and NSF (USA).
Individuals have received support from the Marie-Curie IEF program (European Union); the Leventis Foundation; the A. P. Sloan Foundation; and
the Alexander von Humboldt Foundation

\bibliography{auto_generated}   % will be created by the tdr script.

%------------------------------------------------------------------------------

\cleardoublepage\appendix\section{The CMS Collaboration \label{app:collab}}\begin{sloppypar}\hyphenpenalty=500\textbf{Yerevan Physics Institute,  Yerevan,  Armenia}\\*[0pt]
S.~Chatrchyan, V.~Khachatryan, A.M.~Sirunyan
\vskip\cmsinstskip
\textbf{Institut f\"{u}r Hochenergiephysik der OeAW,  Wien,  Austria}\\*[0pt]
W.~Adam, B.~Arnold, H.~Bergauer, T.~Bergauer, M.~Dragicevic, M.~Eichberger, J.~Er\"{o}
% ...
\vskip\cmsinstskip
% ...

\end{sloppypar}
\end{document}